\documentclass[lettersize,journal]{IEEEtran}
\usepackage{amsmath,amsfonts}
\usepackage{algorithmic}
\usepackage{algorithm}
\usepackage{array}
\usepackage[caption=false,font=normalsize,labelfont=sf,textfont=sf]{subfig}
\usepackage{textcomp}
\usepackage{stfloats}
\usepackage{url}
\usepackage{verbatim}
\usepackage{graphicx}
\usepackage{cite}
\usepackage{color}
\begin{document}

\title{A Secure Semantic Communication System Based on Knowledge Graph}

\author{Qin Guo, Haonan Tong, Sihua Wang,~\IEEEmembership{Member,~IEEE}, Peiyuan Si,~\IEEEmembership{Student Member,~IEEE},
\\ Jun Zhao,~\IEEEmembership{Member,~IEEE}, and Changchuan Yin,~\IEEEmembership{Senior Member,~IEEE}
\thanks{ Q. Guo, S. Wang, and C. Yin are with the Beijing Key Laboratory of
Network System Architecture and Convergence, and also with the Beijing
Advanced Information Network Laboratory, Beijing University of Posts and
Telecommunications, Beijing, 100876 China: Emails: 2785656925@bupt.edu.cn,
 shwang@bupt.edu.cn,  ccyin@bupt.edu.cn.

H. Tong is with the Key Laboratory of Target Cognition and Application Technology (TCAT), Aerospace Information Research Institute, Chinese Academy of Sciences, Beijing 100190, China.
E-mail:tonghn@aircas.ac.cn

P. Si and J. Zhao are with College of Computing and Data Science~(CCDS), Nanyang Technological University~(NTU), Singapore. 
E-mail: peiyuan001@e.ntu.edu.sg,  junzhao@ntu.edu.sg.

This work was supported by Beijing Natural Science Foundation under Grant L223027, the National Natural Science Foundation of China under Grant 62471056 and Grant 62331027, and in part by the China 973 Program under Grant 2009CB320407.
}

}


\IEEEpubid{ }

\maketitle

\begin{abstract}
This study proposes a novel approach to ensure the security of textual data transmission in a semantic communication system. In the proposed system, a sender transmits textual information to a receiver, while a potential eavesdropper attempts to intercept the information.  At the sender side, the text is initially preprocessed, where each sentence is annotated with its corresponding topic, and subsequently extracted into a knowledge graph. To achieve the secure transmission of the knowledge graph, we propose a channel encryption scheme that integrates constellation diagonal transformation with multi-parameter weighted fractional Fourier transform (MP-WFRFT). At the receiver side, the textual data is first decrypted, and then recovered via a transformer model. Experimental results demonstrate that the proposed method reduces the probability of information compromise. The legitimate receiver achieves a Bilingual Evaluation Understudy (BLEU) score of 0.9, whereas the  BLEU score of the eavesdropper remains below 0.3. Compared to the baselines, the proposed method can improve the security  by up to  $20\%$.
\end{abstract}


\section{Introduction}

\IEEEPARstart{W}{ith} the rapid development of technologies such as extended reality (XR), perceptual interconnectivity, and digital twins, the volume of data transmitted over wireless networks has surged dramatically \cite{10211567,10906634}. To address this challenge, one feasible method is to reduce the amount of transmitted data by using source coding techniques, including Huffman coding, arithmetic coding, run-length encoding (RLE), and Lempel-Ziv-Welch (LZW) algorithms \cite{9105909}. However, traditional methods have reached their capacity limits, as increasing transmission dimensions and altering resource usage have rendered them insufficient to meet escalating demands. In contrast, semantic information transmission, which utilizes shared knowledge and semantic understanding between the sender and receiver, presents a promising alternative for significantly reducing the volume of data transmission \cite{10634888,9685654,10740049}.

Due to the openness of wireless channels, semantic communication systems are vulnerable to various security threats \cite{10292907}. Attackers can target both the semantic and physical layers, employing tactics such as backdoor attacks, data poisoning, model theft, privacy breaches, and eavesdropping. Given the risks posed by these attacks, we focus on two key aspects: model security, which addresses attacks that undermine the integrity of the model; and data security, which pertains to the interception of data, particularly through unauthorized eavesdropping.

  Several works in the literature have explored model security in semantic communications. In \cite{yang2023adversarial}, researchers investigated adversarial attacks on semantic communication systems and proposed a novel defense mechanism that enhances model robustness by generating balanced adversarial samples. This approach effectively mitigates malicious attacks, ensuring the stability and reliability of the communication system. In \cite{10478651}, a new method combining deep neural networks (NNs) with differential privacy techniques was introduced. By adding noise to the data during model training, the proposed method protects individual information, significantly enhancing the security of sensitive data in semantic communication systems. In \cite{9979702}, a semantic extraction model training scheme based on federated learning was proposed  which enables distributed learning while incorporating data privacy protection, further enhancing system security. Despite these advancements, several challenges remain unresolved, particularly concerning model complexity, generalizability, and the additional computational resources required.
  
 Another important security issue in the context of semantic communication system  is data security, which forms the central theme of this paper. In \cite{iyer2023survey}, a semantic security communication method based on hashing is proposed which leverages supervised learning to generate a unique, one-time signature or hash code for each message, thereby enhancing security by preventing forgery and tampering. However, a key limitation of hash algorithms is that their fixed output length constrains the potential range of hash values. As the data size increases, the probability of collisions rises, which, in turn, undermines the overall security of the system. In \cite{10107616}, the authors proposed an encrypted semantic communication system for privacy protection, which incorporates an NN-based adversarial encryption training scheme. This system introduces potential attack patterns and performs well on specific datasets; however, it lacks generalization capabilities. Semantic communication based on NN outperforms traditional communication methods in terms of semantic metrics, particularly in low signal-to-noise ratio (SNR) scenarios. However, NN-based encryption schemes suffer from limited interpretability, making it difficult to ensure controllable encryption and decryption processes. The study presented in \cite{10278612} initially proposed a joint source-channel coding scheme for wireless image transmission,  
  where end-to-end data security is achieved by introducing random errors during the encryption process. However, these random errors may also introduce distortions into the decrypted data, potentially compromising data accuracy.
In \cite{xue2022structured}, the authors focus on the application of symmetric searchable encryption to the knowledge graph, aiming to facilitate efficient and privacy-preserving queries over the data while ensuring the protection of sensitive information. However, the proposed scheme suffers from low computational efficiency and significant communication overhead. In \cite{xue2022structured}, the low computational efficiency of the proposed solution is mainly attributed to three major constraints. First, the high computational scale resulting from the construction of encrypted index structures. Second, the significant storage overhead due to encrypted indexes, matrices, and attribute graphs with numerous attribute-key pairs, which drastically increase storage costs. Third, the considerable overhead in associative operations, as complex queries require the combination of multiple encryption operations, further escalates both computational and communication overhead.

Data encryption is a method for protecting semantic information, preventing attackers from reconstructing the original content from intercepted data. 
However, current data encryption schemes face several challenges, including poor interpretability, limited generalization, and decryption errors caused by random noise in \cite{iyer2023survey, 10278612, 10107616,xue2022structured}.

\begin{table}[!t]
\caption{LIST OF MAIN NOTATION.\label{tab:table1}}
\centering
\begin{tabular}{|c|m{0.7\linewidth}|}
\hline
\textbf{Notation} & \textbf{Description} \\ 
\hline
$A$   & Text data \\
\hline
${A}' $   & Topic data \\
\hline
$\tilde{A}$   & Clustering data \\ 
\hline
${E_{vj}}$   & Expansion node \\ 
\hline
$\mu ({v_i},{E_{vj}})$   & Entity relationship \\
\hline
$G$  & Constructed knowledge graph \\ 
\hline
$D^{\prime}$ & Constellation after diagonal encryption of Semantic information \\
\hline
$\varphi_{k}$ & The topic distribution of the POI vocabulary \\ 
\hline
$\theta _{d}$ & The topic distribution of the document \\
\hline 
$E^{ \prime\prime}$ & Semantic information after Fourier inversion with multiple parameters \\
\hline
$E^{ \prime}$ & Semantic information after constellation rotation decryption \\
\hline
$D^{\prime \prime}$ & Semantic information after multiparameter Fourier transform \\
\hline 
$\hat{G}$ &Decrypted data\\
\hline 
$Z$ &Transmitted symbol\\
\hline 
$\hat{Z}$ & Received data by the receiver after being transmitted over the wireless channel\\
\hline 
$\hat{A}$ &Recovered text\\
\hline
\end{tabular}
\end{table}
To this end, we propose a novel encryption method for text semantic communication based on knowledge graphs. This scheme integrates constellation diagonal encryption with multi-parameter weighted Fourier transform to achieve efficient and lightweight data encryption. By employing fundamental operations such as amplitude scaling, phase rotation, and complex multiplication, the framework eliminates the need for constructing complex indexing structures and frequent encryption reconfigurations, thus significantly reducing the computational scale. Furthermore, by leveraging knowledge graphs to extract entities and relationships from text, the scheme ensures enhanced interpretability. The main contributions of this paper are summarized as follows:

(1) We propose a novel secure semantic communication system based on knowledge graphs, which extracts semantic features from the graph to reduce data transmission redundancy and enhance system security. Additionally, the semantic information derived from the knowledge graph represents entities and their relationships, providing greater interpretability.

(2) To address the challenge of limited generalization in encryption schemes, we propose, for the first time, a hybrid encryption approach that leverages knowledge graphs and focuses on encrypting signals in the frequency domain.

(3) We propose using the Detection Failure Probability (DFP) as a metric to assess the security performance for our proposed system. The DFP measures the likelihood that an eavesdropper fails to detect transmission activity over multiple attempts during the data transmission period. A higher DFP value indicates stronger security for wireless communication. By evaluating DFP, the effectiveness of the semantic security approach in protecting wireless communication data can be validated.

(4) Simulation results demonstrate that our proposed method significantly improves the security performance of semantic communication systems while enabling high-quality reconstruction of textual information.

The remainder of this paper is organized as follows. The system model based on a knowledge graph is formulated in Section II. Each functional module of the knowledge graph-based semantic security communication system is introduced in Section III. In Section IV, simulation results on security are presented and discussed. The conclusions are drawn in Section V. 

\section{System Model}

\begin{figure*}[t]
\centering
\includegraphics[width = 7.2in]{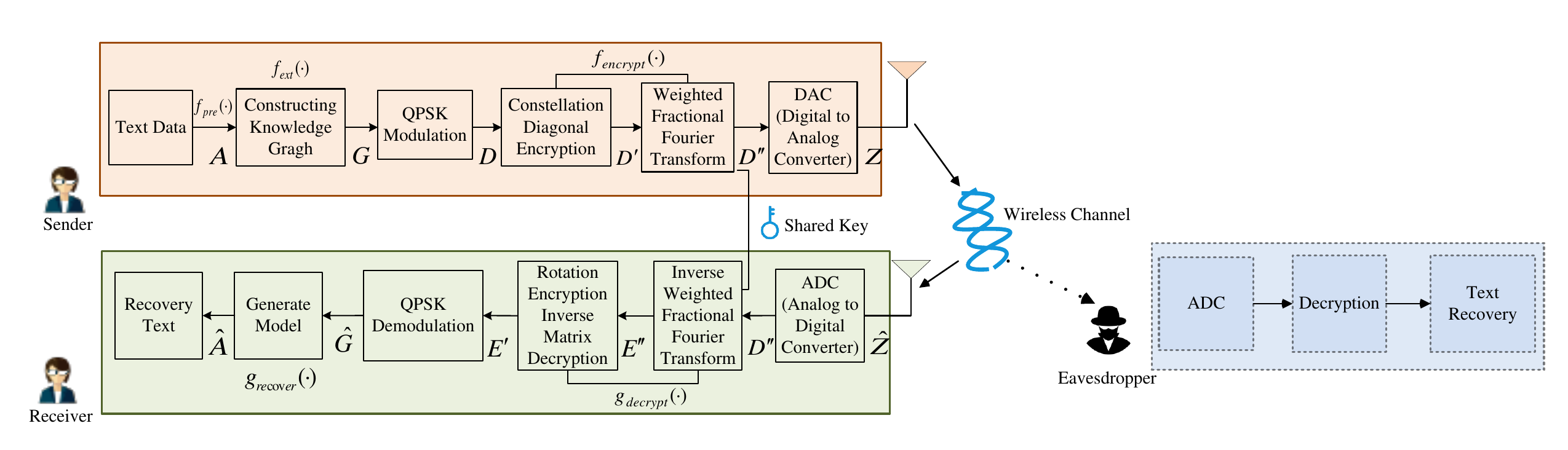}
\caption{The architecture of the proposed secure semantic communication system.}
\label{fig_1}
\end{figure*}

Consider a wireless text information communication system, as illustrated in Fig.~\ref{fig_1}, consisting of a sender, a receiver, and an eavesdropper which attempts to intercept the transmitted signals. Specifically, the sender is equipped with two distinct modules: a semantic extraction module and an encryption module. Simultaneously, the receiver is equipped with a decryption module and a semantic recovery module. To ensure the secure transmission of data, the objective of the system design is to maximize the accuracy of text information received by legitimate users, while minimizing the accuracy of text information recovered by eavesdroppers. To achieve this, the sender needs to securely transmit the textual semantics to the receiver while preventing the eavesdroppers from intercepting the transmitted semantic information. In this context, the semantic extraction module on the sender side utilizes a knowledge graph to refine the key textual semantics. The encryption module employs a hybrid encryption approach, combining constellation diagonal encryption \cite{9133108} with a multi-parameter weighted Fourier transform, to facilitate secure data transmission. Correspondingly, at the receiver, the decryption module extracts the semantic information (i.e., the knowledge graph) from the received encrypted data, and the semantic recovery module is used to convert the knowledge graph into the original text data. The eavesdropper intercepts the transmitted information, which is initially subjected to decryption. After the combined encryption has been decrypted, the text recovery process ensues. However, due to the absence of a decryption key and the lack of a shared knowledge base with the sender, it is extremely difficult for the eavesdropper to reconstruct the original text.

 We denote the system input text data as $A$, the data $A$ is first preprocessed using Latent Dirichlet Allocation (LDA) model \cite{10552275}, followed by the K-means clustering algorithm \cite{10596135} for topic clustering, given by
\begin{equation} 
\tilde{A}=f_{pre}(A),
\end{equation}
where \( f_{pre}(\cdot) \) 
 represents a data preprocessing function, which will be introduced in Section \ref{Text Preprocessing}.
After the data preprocessing, the entities and relationships in the text are extracted into a knowledge graph $G$ via a semantic extraction function, which can be represented as
\begin{equation} 
G=f_{ext} (\tilde{A}),
\end{equation}
where \( f_{ext}(\cdot) \) represents the semantic extraction function, which will be introduced in Section \ref{Semantic Knowledge Graph Construction}.
The knowledge graph is initially modulated using  Quadrature Phase Shift Keying (QPSK), followed by channel encryption and the digital-to-analog conversion module to obtain channel symbol $Z$, which is subsequently transmitted through the wireless channel. $Z$ is represented as
\begin{equation} 
Z=f_{encrypt} (G),
\end{equation}
where  
\( f_{encrypt}(\cdot) \) represents the encryption function, which will be introduced in Section \ref{Channel Encryption}. 

Consider two impacts that are commonly experienced in communication systems: the Additive White Gaussian Noise (AWGN) channel and the slow fading channel. In the AWGN channel, the transmission function is defined as $\eta_n(z) = z + \boldsymbol {n}$, where $z$ represents the input signal and $\boldsymbol{n}$ denotes the additive noise. The noise vector $n$ resides in the complex domain $\mathbb{C}^k$ and consists of independently and identically distributed (i.i.d.) samples, which follow a circularly symmetric complex Gaussian distribution, denoted as $n \sim \mathcal{CN}(0, \sigma^2 I_k)$. Here, $\sigma^2$ represents the variance of the noise, and $I_k$ represents the $k \times k$ identity matrix. In contrast, within the slow fading channel, the transmission function is expressed as $\eta_h(z) = hz$, where $h$ represents the channel gain, which imparts a multiplicative effect on the transmitted signal. The channel gain $h$ follows a complex Gaussian distribution, described by $h \sim \mathcal{CN}(0, H_c)$, with $H_c$ denoting the variance of the channel gain. After the semantic information is transmitted through the physical channel, the received signal is given by
\begin{equation} 
\hat{Z} =hZ+ \boldsymbol{n},
\end{equation}
where $h$ represents the channel gain \cite{he2022representation}, $\boldsymbol {n}$ is the superimposed
Gaussian white noise signal with zero mean and variance of $\sigma _{n} ^{2} $.

At the receiver, the received signal is processed to obtain the decrypted Knowledge graph $\hat{G}$, which is given by
\begin{equation} 
\hat{G} =g_{decrypt} (\hat{Z} ),
\end{equation}
where 
$ g_{decrypt}(\cdot)$ represents the decryption function, which will be introduced in Section \ref{Channel Decryption}. Based on the shared knowledge base (KB) between the sender and receiver, the decrypted and demodulated knowledge graph is converted into the recovered text 
$\hat{A}$ using a Transformer-based generative model \cite{10258000}. The recovery process is given by
\begin{equation} 
\hat{A}=g_{recover}{(\hat{G })=f_{{recover}} f_{{decrypt}}(A)},
\end{equation}
where $ g_{recover}(\cdot) $ represents the text recovery function, which will be introduced in Section \ref{Text Recovery from Semantics}. $f_{{decrypt}^{-1} (\cdot)}$   denotes the decryption function at the receiver's side, which corresponds to the function used at the sender's side. Similarly, $f_{{recover}}^{-1}(\cdot) $  refers to the text recovery function at the receiver's end, serving as the inverse of the corresponding function used at the sender's side. 

The design aim of the entire communication system is to maximize the semantic accuracy of text recovery for legitimate users while minimizing the semantic accuracy of text recovery  for eavesdroppers, which can be given by
\begin{equation} 
\max\left\{BLEU({A}, \hat{A}^h) - BLEU({A}, \hat{A}^f)\right\},
\end{equation}
where 
$ BLEU(A^h, \hat{A}^h)$ represents the accuracy of text recovery for legitimate users, while 
$ BLEU(A^f, \hat{A}^f)$ denotes the accuracy of text recovery for eavesdroppers.
The BLEU score is calculated by counting the frequencies of matching \textit{n}-grams between two sentences \cite{zhou2023comparative}, given by
\begin{equation} 
BLEU = \exp\left\{\min\left(1 - \frac{{l}_{{\hat{A}}}}{{l}_{A}}, 0\right) + \sum_{n} u_n \log P_n\right\},
\end{equation}
where ${{l}_{A}}$ and ${{l}_{{\hat{A}}}}$ represent the lengths of the source text data $A$ and the received text data $\hat{A}$, respectively, i.e., the number of words they contain. $u_n$ denotes the weight of the \textit{n}-grams, and $P_n = \frac{\sum_\kappa \min(C_\kappa(\hat{A}), C_\kappa(A))}{\sum_\kappa \min(C_\kappa(\hat{A}))}
$ represents the \textit{n}-grams score,
 $C_\kappa(\cdot)$ represents the frequency count function for the $\kappa$-th element in the \textit{n}-grams.

In the following, we will describe the computational process for each module. The main symbols involved in the paper are summarized in Table~\ref{tab:table1}.

\section{Knowledge Graph and System Implementation}
In this section, we first describe the process of constructing a knowledge graph from the original text. We then explain how the graph is encrypted for transmission over a wireless channel. Finally, we outline the decryption process at the receiver side to recover the original text.

\subsection{ Knowledge Graph Construction and Encryption at the Sender}
\begin{figure}[H]
\centering
\includegraphics[width=0.48\textwidth]{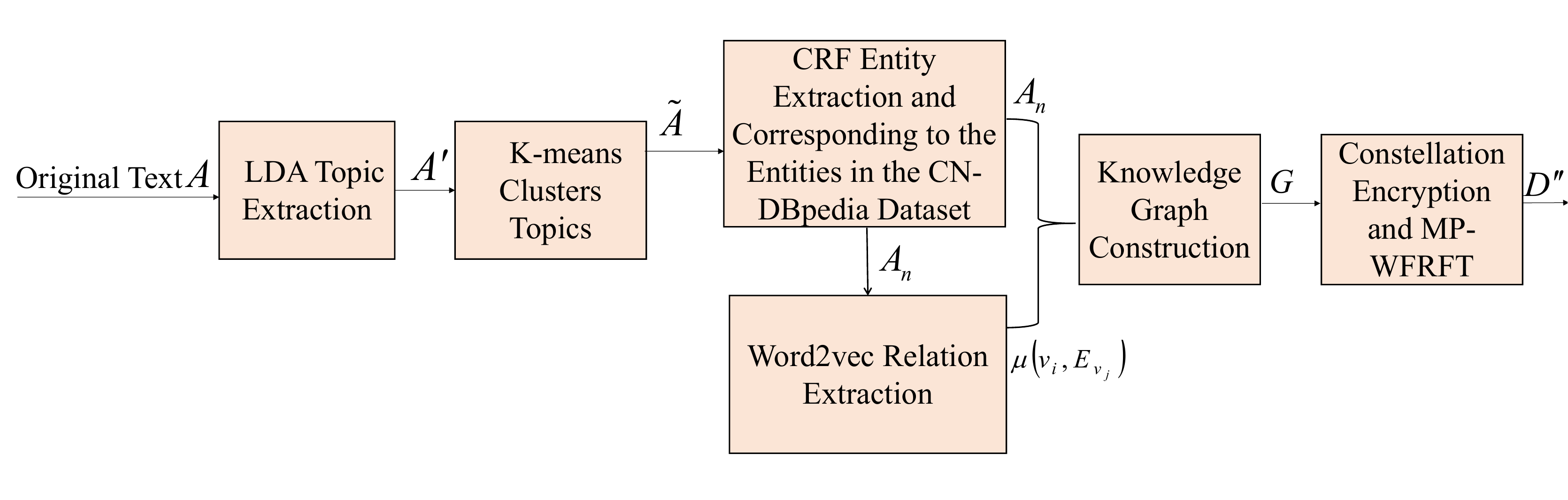}

\caption{Text-based knowledge graph construction     and encryption Processing.}
\label{Fig83}
\end{figure}

At the transmitter, textual data undergoes a multi-stage secure encryption process comprising LDA topic modeling, K-means clustering, CRF entity extraction, Word2Vec embedding, Constellation encryption, and MP-WFRFT encryption. To enhance inter-module coherence, Fig.~\ref{Fig83} comprehensively details the text encryption workflow.

\subsubsection {\label{Text Preprocessing}\textbf{Introduce  $ f_{pre}(\cdot) $ for Text Preprocessing}}
\begin{figure}[]
\centering
\includegraphics[width=0.48\textwidth]{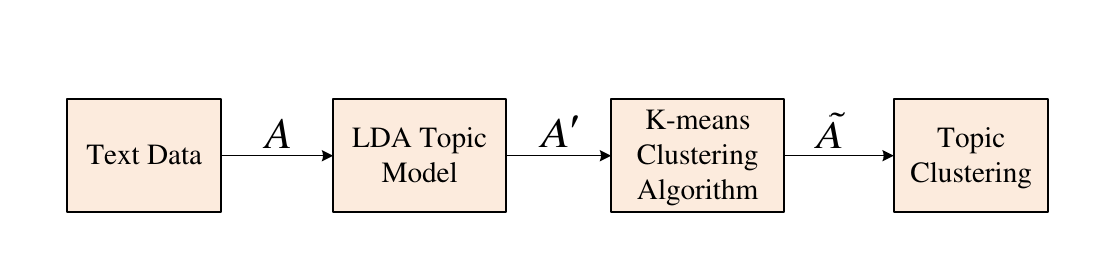}
\caption{Flow diagram of data preprocessing.\label{fig2}}
\end{figure}

 We utilize the LDA topic model and the K-means clustering algorithm for preprocessing the text, as illustrated in Fig.~\ref{fig2}. The LDA model is a generative probabilistic model that assumes each document is generated from a probability distribution over topics, with each topic being derived from a probability distribution over words \cite{10603977,10256673,mohebbi2022computing}. The LDA model calculates the topic distribution for a document and the word distribution for each topic using statistical methods. The probability distributions over topics are derived from the distribution over the document's Points of Interest (POI), which represent the topics of focus. The probability distribution of the POI is determined by the topic distribution of the POI vocabulary, denoted as $\varphi_{k}$, and the topic distribution of the document, denoted as 
$\theta_{d}$.

\textbf{Initialization:} Each word is stochastically assigned a topic $z_n$.

\textbf{Iterative Update:} In each iteration, for each word  $ w_n$, the conditional probability of the word belonging to each topic is computed, and the topic assignment is updated based on the probability $p(w_n|z_k) = \frac{n_{w_k} + \beta}{n_k + N\beta}$, where $n_{w_k} $ denotes the number of occurrences of word $ w_n$  in topic $ z_k$, and $ n_k $ represents the total number of words assigned to topic $ z_k$. The  conditional probability of a word $ w_n$ being assigned to topic $z_k$ at iteration $t$ is given by
\begin{equation}
 P(z_n=z_k\mid  w_n, \theta _{d},\varphi_{k}  ) \propto \frac{n_{d, k}^{(t-1)}+\phi }{n_d^{(t-1)}+K\phi } \cdot \frac{n_{k, w_n}^{(t-1)}+\beta}{n_k^{(t-1)}+N\beta}, 
 \end{equation}
 where $n_{d, k}^{(t-1)}$ represents the number of words assigned to topic $z_k$ in document $d$ at iteration $t-1$, $n_{k, w_n}^{(t-1)}$ represents the number of occurrences of word $w_n$ in topic $z_k$ at iteration $t-1$, $n_d^{(t-1)}$ denotes the total word count in document  $d$ at iteration $t-1$, and $n_k^{(t-1)}$ represents the total number of words assigned to topic $z_k$ at iteration $t-1$. Additionally, $N$ is the size of the vocabulary, $K$ denotes the total number of topics, and $\phi $ is a hyperparameter controlling the sparsity of the topic distribution within each document. Finally, $\beta$ is a smoothing hyperparameter that ensures that the frequency of a word is nonzero in a given topic.

After assigning topics to each word $w_n$ in document $d$ using the LDA topic model, we invoke Kullback-Leibler $( KL )$ divergence as a measure of semantic similarity, denoted $D_{\_s}$, to quantify the difference between the probability distributions of two topics \cite{wang2023generalized}.

\begin{equation}
\begin{aligned}
D_{\_s}&=J_{s} \left ( \theta _{d1} \parallel \theta _{d2}  \right ),\\
     &=\frac{1}{2} KL\left ( \theta _{d1} \parallel \frac{\theta _{d1} +  
\theta _{d2}}{2} \right )+  \frac{1}{2} KL\left ( \theta _{d2}  \parallel \frac{\theta _{d1} +  \theta _{d2} }{2} \right ), 
\end{aligned}
\end{equation}
where $\theta _{d1} $  and $\theta _{d2} $  are the topic probability distributions of the document. $ KL $ Divergence measures the difference between two probability distributions, which is given by
$KL\left ( \theta _{d1} \parallel \theta _{d2} \right ) =\sum_{\varsigma }\theta _{d1} \left ( \varsigma   \right )log\left ( \frac{\theta _{d1} \left ( \varsigma   \right ) }{\theta _{d2} \left ( \varsigma  \right ) }  \right )  $, $\theta _{d} \sim Dirichlet\left ( \varsigma  \right ) $, where $\varsigma $ is the hyperparameter of the Dirichlet distribution, which controls the sparsity of the topic distribution\cite{uthirapathy2023topic,10038668,9723527}. Given that $ D_{\_s} $ can be revealed, each word is assigned to a topic from a finite set, and similar topics can be grouped into a cluster, thereby providing the foundation for clustering.

Next, topic clustering is performed to group highly related words into distinct topics. We utilize the computationally efficient and fast-converging K-means clustering algorithm to cluster words with similar content into the same group, with each group representing a distinct topic, thereby achieving topic clustering. Given a vocabulary of size $N$, 
where each word is represented as a $d$-dimensional vector, the goal is to cluster these data points into $K$  distinct clusters. The process begins by randomly selecting $K$ initial centroids $\mu_1, \mu_2, \ldots, \mu_K$. For each vocabulary  $w_n$, we compute its distance to each centroid and assign it to the closest centroid.
\begin{equation}
c_n = \arg\min_K \| w_n - \mu_K \|,
\end{equation}
where $c_n$ represents the index of the cluster to which the vocabulary item $w_n$ is assigned.

Update the centroid for each cluster:
\begin{equation}
\mu_K = \frac{1}{|C_K|} \sum_{w_n \in C_K} w_n,
\end{equation}
where $C_K$ denotes the set of all data points in the $K$-th cluster, and $|C_K|$ represents the total number of data points in that cluster.

Repeat the above steps until the clustering results converge, thereby completing the topic clustering process.

In the data preprocessing phase, we introduce two processing steps involving LDA and K-means. During the training phase, LDA is employed to learn a fixed topic distribution from the corpus, estimating topic-word and document-topic distributions. Once the model is trained, these parameters remain constant and are not subject to further updates. In the inference phase, LDA infers topic proportions for new documents based on the pre-trained model, without any dynamic updates. Subsequently, K-means clustering is applied to these inferred topic distributions (or synthetic data generated from them), but K-means is not utilized during the training of the LDA model. Notably, the LDA model remains static during inference, with topic distributions for new documents derived from the fixed parameters learned during the training phase. This approach avoids dynamic updates, ensuring that the LDA model remains unchanged during testing. Additionally, synthetic data generated from the topic distributions ($\theta_{d}$ and $\phi_{k}$) learned by LDA are used as input to the K-means clustering algorithm, ensuring that the clustering results reflect the latent topic structure captured by the model.

\subsubsection  {\label{Semantic Knowledge Graph Construction}\textbf{Introduce  $ f_{ext}(\cdot) $ for Constructing Semantic Knowledge Graph}}
\begin{figure}[!t]
\centering
\includegraphics[width=0.48\textwidth]{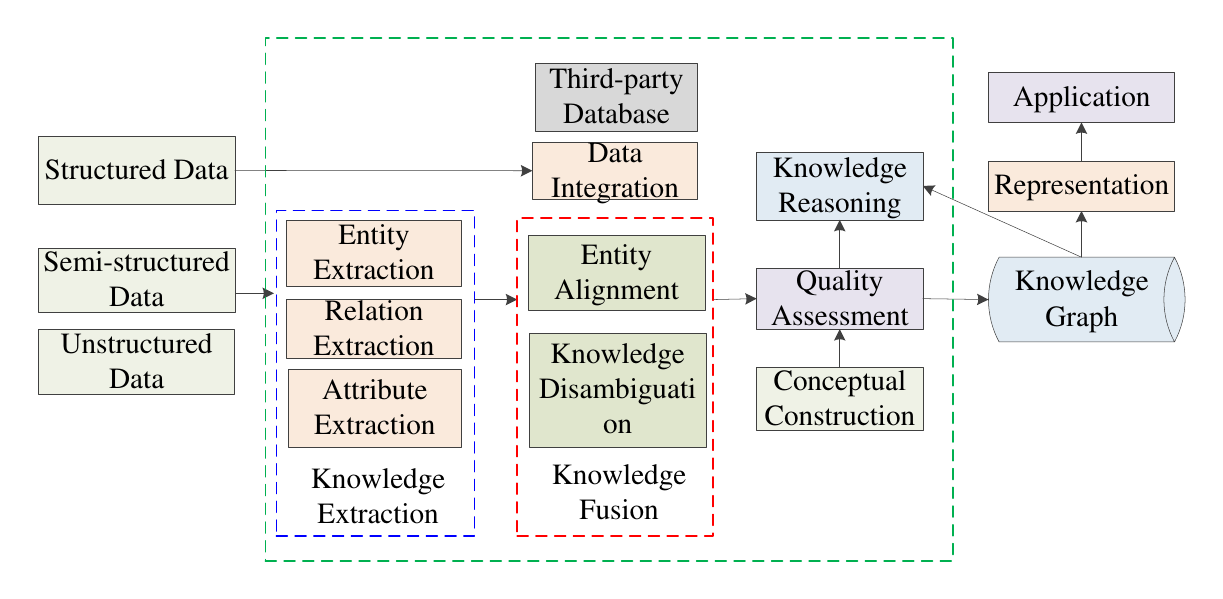}
\centering
\caption{Architecture of knowledge graph construction.\label{fig3}}
\end{figure}

\begin{figure}[!t]
\centering
\includegraphics[width=0.48\textwidth]{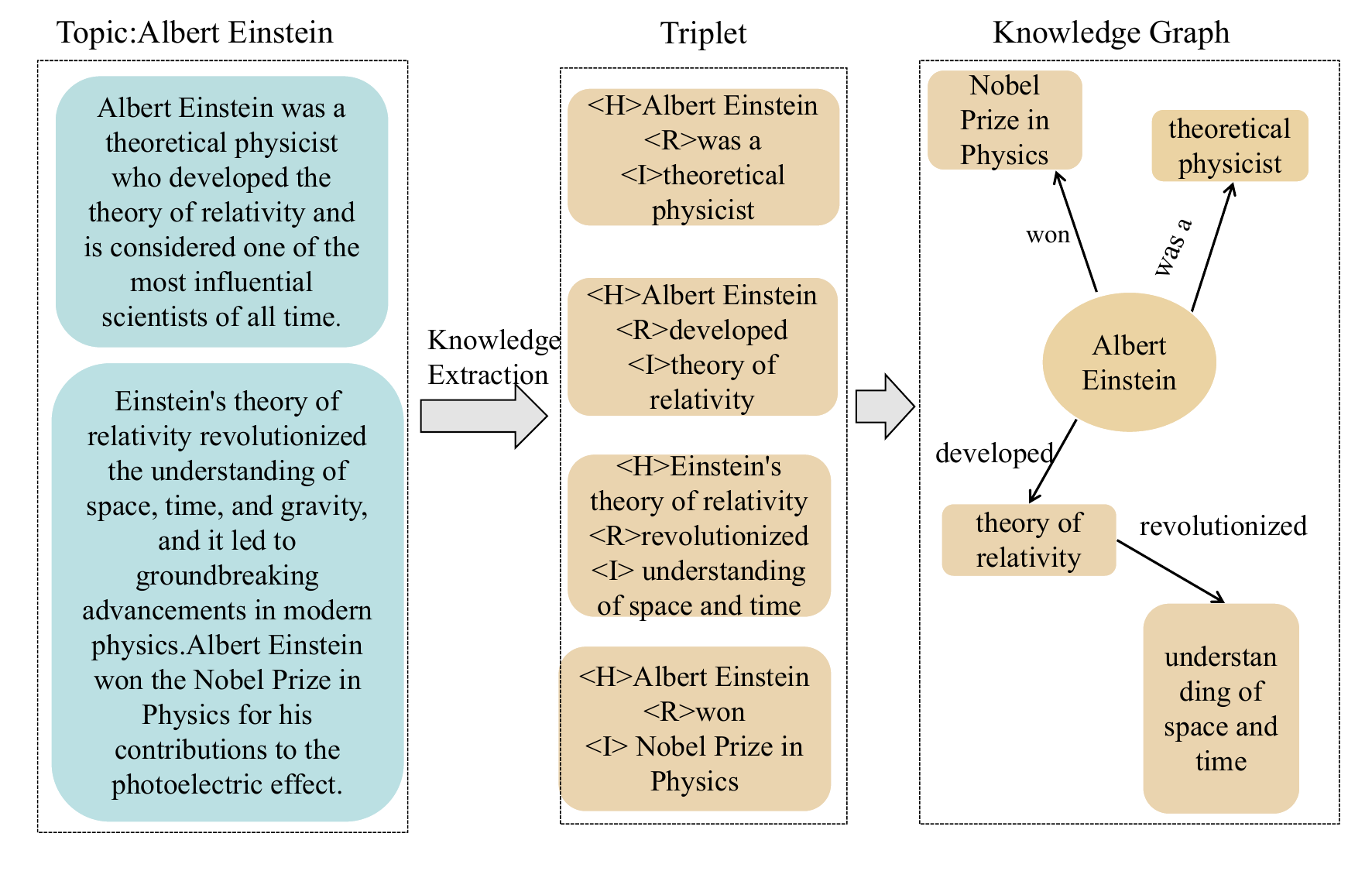}
\centering
\caption{A visualization example of knowledge graph.\label{Fig13}}
\end{figure}

After performing text data clustering, a knowledge graph can be constructed using entity recognition and relation extraction techniques applied to the clustered data \cite{liang2023hyper, gao2022detecting, srivastava2023robust}. The architecture and visualization examples of the knowledge graph are illustrated in Fig.~\ref{fig3} and Fig.~\ref{Fig13}. Fig.~\ref{fig3} illustrates the framework of the knowledge graph, which encompasses knowledge extraction, fusion, and application. Meanwhile, Fig.~\ref{Fig13} presents a visual example of constructing a knowledge graph from textual data. we will provide detailed descriptions
of the key components in Fig.~\ref{fig3} to enhance clarity. The construction of knowledge graphs is a systematic process that starts with the classification and processing of multi-source heterogeneous data: structured data (e.g., relational databases), semi-structured data (e.g., JSON/XML), and unstructured data (e.g., text, images). Key techniques like entity extraction, relation extraction, attribute extraction, and knowledge extraction are used to transform raw data into structured information. Third-party knowledge bases (e.g., Wikidata) are integrated to enhance coverage, followed by entity alignment and conflict resolution. Knowledge refinement is achieved through disambiguation, fusion, and reasoning, culminating in quality assessment to ensure accuracy, completeness, and consistency. This framework supports the creation of an interconnected knowledge base.

\begin{figure}[!t]
\centering
\includegraphics[width=0.48\textwidth]{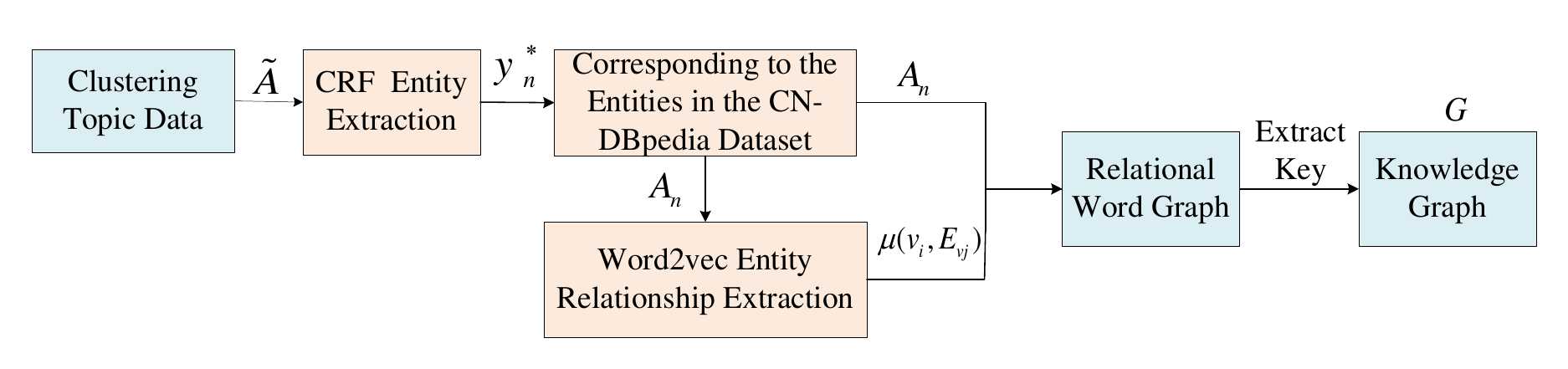}
\centering
\caption{Knowledge graph construction process.\label{fig4}}
\end{figure}

To extract entities from each clustering topic 
$\tilde{A}$ , the Conditional Random Field (CRF) is utilized. CRF is a well-established and efficient method for obtaining entities. This approach functions by performing pattern matching, leveraging the input sentence and predefined rules to identify and extract entities. As a discriminative probabilistic model, CRF differs from generative models in modeling the conditional probability of output labels given input features, rather than modeling the joint probability of both input and output. Subsequently, the extracted entities are mapped to their corresponding entities in the CN-DBpedia dataset \cite{10184660}.  Word2Vec \cite{10215867} is a widely recognized technique to learn distributed word representations (word embeddings) from large corpora of text data. It is grounded in the principle that words occurring in similar contexts tend to have similar meanings. The method embeds words into continuous vector spaces, where semantically similar words are positioned closer together. Subsequently, Word2Vec is employed to capture the relationships between these entities, thereby constructing a semantic relation word graph. PageRank \cite{9888678} is utilized to compute the ranking scores of semantically significant terms, aiding in the identification of key keywords and ultimately contributing to the construction of a knowledge graph. The framework of the knowledge graph is illustrated in Fig.~\ref{fig4}.
 
 By directly modeling the conditional probability of a given observation sequence, CRF can more accurately capture the overall structure and contextual information of a sentence. It then predicts and outputs the most probable labeling sequence based on the target observation sequence \cite{LSTM-CRF-drug-ner}. The formula for the CRF model is given by
\begin{equation}
p(y_{n}  \mid w_{n} ,\boldsymbol {v }  ) = \frac{\exp\left(\boldsymbol {v }^T \varphi(w_{n}, y_{n})\right)}{Z(\boldsymbol {v }, w_{n})},
\end{equation}
where $w_n$ represents the input derived from the cluster topic data $\tilde{A} = \left( \widehat{z}_{1} , \widehat{z}_{2} , \dots, \widehat{z}_{m}  \right)$, $y_n$ denotes the predicted label, $\varphi(w_n, y_n)$ is the feature vector, $\boldsymbol{v}$ is the parameter vector, and $Z(\boldsymbol{v}, w_n)$ represents the normalization factor, summing probabilities over all possible values of $y_n$. The predicted label $y_n^*$ is obtained by leveraging the input sequence $w_n$ and the parameter vector $\boldsymbol{v}$, which is given by

\begin{equation}
y_{n}^{*} =arg\underset{y_{n} }{max} P(y_{n} /w_{n} ,\boldsymbol {v }).
\end{equation}

 The entities are first extracted as $y_{n}^{*}$ and then aligned with the corresponding entities in the CN-DBpedia dataset, resulting in ${A_n} = \left\{ v_1, v_2, \dots, v_\eta \right\}$ \cite{li2023deep,zhang2022cross}.
 Following this, the process of extracting entity relationships using Word2Vec is presented in detail.

 The knowledge graph in the CN-DBpedia dataset is denoted as ${G_{kg}} = (V, E)$, where $V$ represents the set of semantic word nodes, and   $E$ represents the set of edges between entities. Any edge corresponds to a triplet, where $E=\left \{ v_{i},v_{j} ,label  \right \}$; ${v_i }$  and  ${v_j} $ represent the starting and ending nodes, respectively, and $lable$  represents the relationship label between the nodes.  ${A_n} $ represents the set of entity nodes, 
$E_v$ represents the set of extended nodes. The cosine similarity between two nodes is computed using Word2Vec to quantify the semantic similarity between $v_i$ and $E_{v_j}$, denoted as $\mu(v_i, E_{v_j})$, as given by

\begin{equation}
\mu(v_i, E_{v_j})=
\frac{\mathbf{v}_{vi} \cdot \mathbf{v}_{Evj}}{\|\mathbf{v}_{vi}| \cdot \|\mathbf{v}_{Evj}\|}.
\end{equation}

\begin{figure}[!t]
\centering
\includegraphics[width=0.4\textwidth]{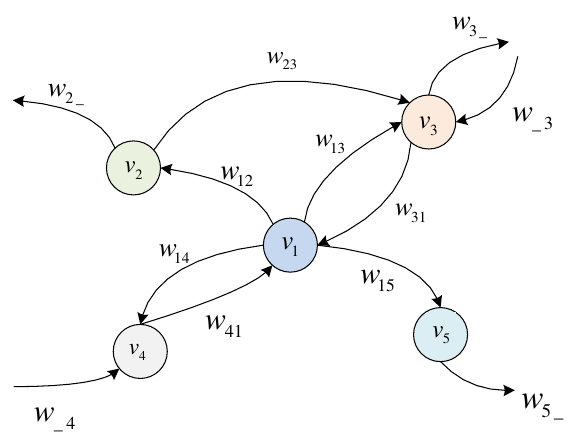}
\centering
\caption{Example of semantic relationship word graph.\label{fig5}}
\end{figure}

The parameter $\mu(v_i, E_{v_j})$ indicates the potential relationships between the entity node $v_i$ and the extended nodes $E_{v_j}$, which are associated with other nodes. To date, by integrating the mapped entities and their relationships, we have constructed a semantic relation word graph. An illustrative example is shown in Fig.~\ref{fig5}.

After constructing the semantic relation word graph, a centrality algorithm is applied to evaluate the significance of the nodes, generating a ranking score for each node. The random transfer probability $ p({v_j}) $ is
\begin{equation}
p({v_j}) = \frac{{f({v_j})}}{{\sum\limits_{j = 1}^\eta {f({v_j})} }},j = 1,2,...,\eta,
\end{equation}
where  $\eta$ represents the number of nodes, and $ f({v_j})$ represents the word frequency of  ${v_j}$. ${W_{ij}}$ represents the jump probability from ${v_i}$ to ${v_j}$, which is given by 
\begin{equation}
{W_{ij}} = \frac{{{T _{eight}}({v_i} \to {v_j})}}{{\sum\limits_\lambda {({T _{eight}}({v_i} \to {v_\lambda}))} }},
\end{equation}
where  ${T _{eight}}({v_i} \to {v_j}) $ represents the weight of  $ {v_i}$ pointing to the $ {v_j}$ side, and $\lambda$ represents all possible entities.

By integrating both edge weights and word frequencies, the PageRank \cite{mobile2023retracted} iteration formula is modified as follows:

\begin{equation}
{P_R}({v_j}) = (1 - a) \times p({v_j}) + a \times \sum\limits_{i \in in(j)} {(T({v_i}) \times {W_{ij}})} , 
\end{equation}
where ${P_R}({v_j})$ denotes the rank score of  ${v_j}$, $p({v_j})$ represents the random transition probability of ${v_j}$, $T({v_i})$ refers to the number of orientation-indicating nodes, $in(j)$ denotes the set of nodes pointing to ${v_j}$, and $a$ is the damping factor used to compute the ranking scores of the nodes.

Finally, PageRank is applied to the semantic relation word graph to compute the ranking scores for the words corresponding to each entity node, extract key semantic elements, and integrate the semantic relation word graph to construct the knowledge graph ${G}$ of the original text.
\subsubsection{\label{Channel Encryption}\textbf{Introduce  $ f_{encrypt}(\cdot)$ for Channel Encryption}}
After the construction of the knowledge graph, the graph $G$ could be intercepted by malicious attackers during transmission, posing a potential security threat. If appropriate security measures are not implemented, the entire transmission could be compromised. Therefore, the use of channel encryption technology is essential to mitigate potential security threats and attacks. 

In this paper, we implement channel encryption using the combination of constellation diagonal encryption and multi-parameter weighted fractional Fourier transform (MP-WFRFT) \cite{10726146}. Next, we will introduce the details of channel encryption. Before the encryption process, modulation is first performed. In this study, the Quadrature Phase Shift Keying (QPSK) modulation technique is employed. QPSK modulation is achieved by altering the phase of the carrier signal, which provides strong resistance to noise and interference, mainly due to its reduced amplitude fluctuations, thereby minimizing the impact of interference on signal quality.
 
To achieve encryption, the constellation points are rotated, with both the amplitude and phase being adjusted to control the encryption process. The rotation key is exclusively held by authorized users, and the legitimate receiver decrypts the message using this key, ensuring secure communication. By employing the correct key, the receiver can recover the original sequence.

Converting the Knowledge Graph 
$G$ into a Semantic Sequence $\boldsymbol{D}$, followed by encryption. Upon encrypting the semantic sequence $ \boldsymbol {D}$, it is first processed in groups, with the length of each group set to $L$. The two-dimensional Ciphertext-Policy Attribute (2D-CPA) chaotic system \cite{8521240} is employed to generate a discrete semantic sequence of length $L$, Initially, the values $ ({f_0},{\zeta _0}) $ are set, where $f_0$ and $\zeta _0$  represent the initial amplitude and the initial phase, respectively. By adjusting the parameters of the chaotic system, the sequence  $ (f,\zeta ) $ 
is derived.


The diagonal amplitude transformation matrix is given by
\begin{equation}
\boldsymbol{F} = diag\left( {\frac{{\left| f \right|}}{{\left| {{f_{\max }}} \right|}} + \varpi } \right),
\end{equation}
where the sequence $f$ represents the amplitude, and the variable $\varpi$ is used to adjust the range of the amplitude transformation.
The diagonal phase rotation matrix $\boldsymbol{\xi}$ is given by
\begin{equation}
 \boldsymbol{\xi}  = diag\left( {{e^{ - 2\pi i\zeta}}} \right),
 \end{equation}
 where $\zeta $ represents the phase rotation factor.

 After obtaining the diagonal amplitude transformation matrix $\boldsymbol{F}$ and the diagonal phase rotation matrix $\boldsymbol{\xi}$, the diagonal encryption matrix $\boldsymbol{W}(f, \zeta)$ for the constellation is constructed using a 2D-CPA chaotic sequence,

\begin{equation}
\boldsymbol{W}(f,\zeta ) =\boldsymbol{F \times \xi}.
\end{equation}

Subsequently, the semantic sequence $\boldsymbol{D}$ is multiplied by the diagonal encryption matrix $\boldsymbol{W}(f, \zeta)$, thereby yielding the ciphertext matrix $\boldsymbol{D'}$,

\begin{equation}
\boldsymbol{D^{\prime}} = \boldsymbol{D} \times \boldsymbol{W}(f,\zeta ).
\end{equation}

By simultaneously encrypting both the amplitude and phase of the original constellation, the constellation distribution undergoes both expansion and rotation, thereby improving the concealment of semantic signals.


The feasibility of constellation diagonal encryption for QPSK signals, implemented through amplitude and phase rotation, has been preliminarily verified from two key perspectives: first, its compatibility with existing QPSK transceiver architectures, facilitating integration into traditional communication systems; second, its relatively low computational complexity, making it suitable for applications in resource-constrained scenarios.
However, despite these advantages, the deployment of this technology in practical systems still faces several potential challenges, primarily including difficulties in carrier synchronization, inaccuracies in channel estimation, and insufficient adaptability in dynamic environments.


The sequence $\boldsymbol{x}$ is obtained by processing $\boldsymbol{D^{\prime}}$. The expression for the multivariate parameter-weighted fractional Fourier transform (MP-WFRFT) \cite{8731955,10281128} of the semantic digital signal sequence $\boldsymbol{x} = [{x_0},{x_1},{x_2},...,{x_{\rho - 1}}]$ is given by

\begin{equation}
f^{\alpha,\boldsymbol{v}}(\boldsymbol{x}) = \sum\limits_{l = 0}^{M - 1} {{\omega _l}} (\alpha,\boldsymbol{v}){F^{4l/M}}(\boldsymbol{x}),
\end{equation}
where the total number of elements in the sequence is represented by $\rho $, $\alpha$ denotes the order of the transformation, the value of $l$ is in the set $\left \{ 0,1,2,3 \right \}$, $M$ represents the variable coefficients, and $\boldsymbol{F}$ is the normalized discrete Fourier transform matrix. 
The scale vector is given
by $ \boldsymbol{v}=[{m_0},{m_1},{m_2},{m_3},{n_0},{n_1},{n_2},{n_3}]$,
 and ${\omega _l}(\alpha ,\boldsymbol{v}) $ represents the set of weighting coefficients. which is given by 
\begin{equation}
\begin{split}
\omega_l(\alpha, \boldsymbol{v}) &= \frac{1}{4} \sum_{k^{'} = 0}^{3} \exp \Bigg( \pm \frac{2\pi i}{4} \Bigg[ (4m_{k^{'}} + 1) \\
& \quad \times\alpha \left( k^{'} + 4n_{k^{'}} \right) - lk^{'} \Bigg] \Bigg).
\end{split}
\end{equation}

Since semantic signals exhibit power invariance and strong signal-hiding characteristics, exemplified by a constellation-like Gaussian distribution in MP-WFRFT without requiring transformation, employing chaotic sequences as control parameters in MP-WFRFT can further enhance the anti-scanning capabilities of its parameters, thereby facilitating high-intensity encryption of the semantics of wireless network communication data.

(1) Determination of Packet Parameter L

In the proposed chaos-based encrypted semantic communication system, the packet number L is determined by both the data dimension and modulation scheme. The fundamental relationship is given by:

\begin{equation}
L=\left [ \left ( M\times N \right ) /X \right ],
\end{equation} 
Where $\boldsymbol{D} \in \mathbb{C}^{M \times N}$ represents the original data matrix, with M denoting the number of data blocks and N indicating complex symbols per block. X specifies the number of QPSK symbols per packet.

Remark 1: For Quadrature Phase Shift Keying (QPSK) modulation, each symbol encodes 2 information bits, resulting in a packet capacity given by:
\begin{equation}
k = 2 \times X \ (\text{bits/packet}),
\end{equation} 
Remark 2: Each packet requires 9 chaotic parameters, specifically $\alpha$ and $\boldsymbol{v}$, for encryption purposes. This results in a total sequence length of 9L, where L is the length of the packet.

(2) Role in Security and Complexity

\textbf{Security Perspective}: A larger value of $L$ enhances the entropy of the chaotic sequence, rendering brute-force attacks computationally impractical. However, when $L$ exceeds $L_{max}$ (where $L_{max}$ represents the saturation threshold of the mixing property of the chaotic attractor), no additional security gains can be achieved.

\textbf{Computational Complexity}: The time complexity of sequence generation is O($9L$), and the design of the $9L$ length can effectively enhance its robustness against known-plaintext attacks, achieving a reasonable balance between complexity and anti-attack performance. From the perspective of parameter utilization, a larger $X$ can improve the utilization rate of chaotic parameters, but it is necessary to take into account the computational complexity (such as the computation load of MP-WFRFT in Equation ($25$), which reflects the trade-off between parameter efficiency and complexity.In addition, the QPSK adopts the design of $k=2X$. While ensuring spectral efficiency, through the orthogonal dimension combination of amplitude, phase and parameter dynamics, it forms a cooperative protection mechanism with chaotic encryption, achieving the collaborative optimization of performance and security under the constraint of complexity.

Subsequently, a discrete semantic sequence $\boldsymbol{\tau} = \left[\tau_1, \tau_2, \dots, \tau_{9L}\right]$ is generated using the 2D-CPA chaotic system. The length of the sequence $\boldsymbol{\tau}$ is nine times the number of signal packets, with each set of nine consecutive values serving as the control parameters required for the MP-WFRFT transformation of the ciphertext matrix $\boldsymbol{D^{\prime}}$. Specifically, $\boldsymbol{D^{\prime}}$ corresponds to the transformation parameter $\alpha = \tau_1$, and $\boldsymbol{v} = [\tau_2, \tau_3, \tau_4, \tau_5, \tau_6, \tau_7, \tau_8, \tau_9]$. The semantic signal undergoes the MP-WFRFT transformation, adds a cyclic prefix, and transitions from parallel to serial transmission before being transmitted through a Gaussian channel after analog-to-digital conversion. After applying the weighted Fourier transform to $\boldsymbol{D^{\prime}}$, the encrypted matrix $\boldsymbol{D^{\prime \prime}}$ is obtained, which is given by 
\begin{equation}
\boldsymbol{D^{\prime \prime}} = \boldsymbol{MP} - f^{\alpha,\boldsymbol{v}}\boldsymbol{(D ^{\prime}, \alpha ,v)},
\end{equation} 
\text{where} $\boldsymbol{M}$ represents the scaling transformation matrix,  and $\boldsymbol{P}$ denotes the translation matrix. 
\begin{equation}
\boldsymbol{M}=\left[ \begin{matrix}
   \omega _{0}^{*} & 0 & 0 & 0  \\
   0 & \omega _{1}^{*} & 0 & 0  \\
   0 & 0 & \omega _{2}^{*} & 0  \\
   0 & 0 & 0 & \omega _{3}^{*}  \\
\end{matrix} \right],
\end{equation} 
where $\omega _{ l}^{*}$ represents the inverse weighted coefficient, $\omega _{ l}^{*} =\omega _{l}(-\alpha ,\boldsymbol{v}) $.
\begin{equation}
\boldsymbol{P}=\left[ \begin{matrix}
  1 & 0 & 0 & 0  \\
   0& 1 & 0 & 0  \\
   0 & 0 &1 & 0  \\
   0 & 0 & 0 &1  \\
\end{matrix} \right],
\end{equation} 
The encrypted matrix $\boldsymbol{D_{\rm{}}^{\prime \prime}}  $  is converted into $Z$ via a Digital-to-Analog Converter (DAC) and subsequently transmitted through the communication channel.

Secure and efficient key management and distribution are essential for ensuring the confidentiality, integrity, and non-repudiation of communications. This study adopts a generic three-tier distributed architecture for key management, aiming to achieve an optimal balance between security and efficiency.
\subsection{Decryption and Text Recovery at the Receiver}
The receiver needs to perform two steps: decryption and recovery. The detailed process is outlined as follows.
\subsubsection{\label{Channel Decryption}\textbf{Introduce  $g_{decrypt}(\cdot)$ for Channel Decryption}}
 At the receiver, the received signals $\hat{Z}$ are converted into  $\boldsymbol{D_{\rm{}}^{\prime \prime}} $. 
 We employ the multi-parameter weighted inverse fractional Fourier transform (MP-WIFFT) and constellation diagonal decryption techniques to decrypt the channel. It obfuscates both the time-domain and frequency-domain characteristics of the signal, relying on rotation matrices, varying weights, and other factors, thereby enhancing the computational complexity of decryption and making it difficult for  eavesdroppers to effectively intercept and decode the information. This hybrid encryption approach enhances the complexity of the decryption process, thereby ensuring the security of semantic information transmission. The semantics encrypted by the chaotic encryption algorithm proposed by this research can be restored through the following steps:

Step 1: Given the private key and varying weight parameters, the MP-WFRFT inverse transformation is performed, which is given by
\begin{equation}
\boldsymbol{E^{\prime\prime}}= \boldsymbol{MP} - f^{\alpha,\boldsymbol{v}}\boldsymbol{(D^{\prime \prime}, - \alpha ,\boldsymbol{v})},
\end{equation}

Step 2: Recover the original communication semantics (i.e., knowledge graph) $\boldsymbol{E^{\prime}}$, which can be given by
\begin{equation}
\boldsymbol{E^{ \prime}} = \boldsymbol{E^{ \prime\prime}} \times \boldsymbol{{W^{ - 1}}}(f,\zeta ),
\end{equation}

After the data is decrypted, the receiver applies QPSK demodulation to accurately recover the phase used in mapping, thereby restoring the processed textual data $\hat{G}$. Subsequently, a generative model is employed to reconstruct the data into its original textual form.

\subsubsection{\label{Text Recovery from Semantics}\textbf{Introduce  $ g_{recover}(\cdot) $ for Text Recovery from Semantics}}
We employ a generative model to recover the text \cite{10525972,10537562}. In text recovery tasks, the process typically involves concatenating the current input vector with the output vector from the previous time step at each decoding step, conditioned on the context labels and the structured information graph embeddings from the knowledge graph $\hat{G}$, to generate the current output. Ultimately, the text sequence with the highest output probability is selected. The formula for the generative model is given by
\begin{equation}
P({X_1},{X_2},...,{X_\rho}) = \prod\limits_{i = 1}^n p ({X_i}\left| {{X_{1:i - 1}},U,L} \right.),
\end{equation}
where ${X_i}$ represents a sequence of word tokens in the knowledge graph $\hat{G} $, and $ p$  represents the maximum probability of the decoding process. The notation $ X_{1:i-1}$, which includes the elements from 1 to $i-1$, refers to the typical decoding input for the current sequence. $U$ represents contextual relevance tokens, such as prior conversation history or document context, while $L$ denotes the structured information graph embedding, which includes syntactic structures, entity relationships within knowledge graphs, and domain-specific rules.

The generative model aims to effectively  produce text content, guided by the auxiliary information $U$ and $L$,
while incorporating dynamic features.

The training objective of the generative model is to maximize the log-likelihood function in (29), which involves optimizing the model's parameters (such as the weights of the neural network) to increase the probability of generating each data point ${X_i}$. The mathematical expression is
\begin{equation}
\hat{\theta} = \arg\max_{\theta} \sum_{i=1}^{n} \log P(X_i; \theta).
\end{equation}

The original loss in the language model is divided into four parts based on 
$U$ and $L$, and joint training is performed using hyperparameter weighting.

\textbf{Encoder  static features:} Splicing $U$ as the input sequence, which is specifically represented as
\begin{equation}
U = u_1^i + u_2^i +  \cdots  + u_\varepsilon ^i,
\end{equation}
where $\varepsilon $  represents the length of the attribute.

Input the sequences $U$ and $L$ into the encoder of the generative model architecture and pass them through the embedding layer to obtain their corresponding vector representations.
The vector representations $\boldsymbol{emb}$ is provided by
\begin{equation}
\boldsymbol{emb} = \left\{ \boldsymbol{emb_{1:m}^U},\boldsymbol{emb_{1:n}^L} \right\}.
\end{equation}

Using the length representation $\boldsymbol{emb_{1:m}^U}$ and $\boldsymbol{emb_{1:n}^L}$ as initial inputs $\boldsymbol{H^i}$ for the encoder side attention network layer, and compute the multi-head attention. The Encoder structure consists of $T$ layers, each containing a self-attention module and a feedforward neural network. The system architecture using a Transformer-based generative model \cite{10391042} is shown in Fig.~\ref{Fig14}. The following calculation formula is as follows:

\begin{figure}[!t]
\centering
\includegraphics[width=0.48\textwidth]{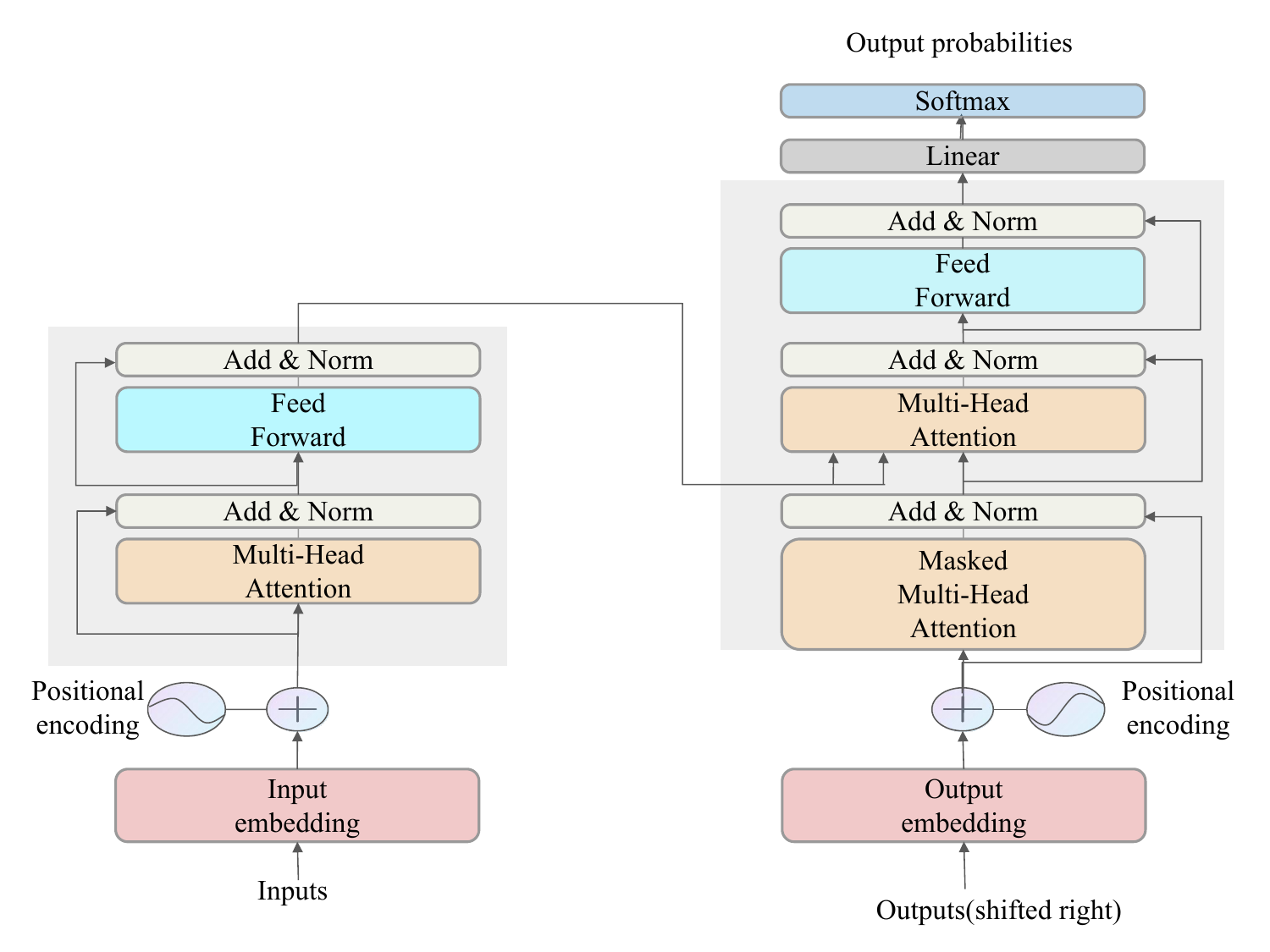}
\centering
\caption{Transformer system architecture.\label{Fig14}}
\end{figure}
\begin{equation}
\boldsymbol{H^{i + 1}} = \mathrm{FNN}(\mathrm{MultiHead}(\boldsymbol{H^i},\boldsymbol{H^i},\boldsymbol{H^i})),
\end{equation}
where $ \mathrm{MultiHead}$ denotes the multi-head attention mechanism, and $\mathrm{FNN}$ refers to a fully connected  feedforward neural network.

\textbf{Decoder fusion attention information:} In the Masked Multi-Head Attention module, the current state $\boldsymbol{H_B}$ from the previous layer of historical information is used as the query vector for static personalized feature representation, denoted as 
$\boldsymbol{Q}$. Let $ \boldsymbol{H^P}$
  represent both the key $K$ and value 
$V$  in the attention module. The attention score can then be calculated as follows:
\begin{equation}
{C} = \mathrm{Softmax}\left( {\frac{1}{{\sqrt d }}(\boldsymbol{H_B}\boldsymbol{W^Q})\boldsymbol{{({H^P}{W^K})}^T}} \right)(\boldsymbol{H^P}\boldsymbol{W^V}),
\end{equation}
\text{where} $\boldsymbol{W^Q}$, $\boldsymbol{W^K }$, and $ \boldsymbol{W^V } $ all represent learnable parameter matrices.

By leveraging self-attention scores, a weighted context vector is generated and passed to the subsequent decoder layer, while the remaining components of the model follow the standard generation architecture. Finally, the decoder module outputs $ \boldsymbol{H^0} $, which is then provided as input to the consistency module.

\textbf{Alignment module consistency generation:} The output $\boldsymbol{Y}$ of the final layer of the decoder module is given by
\begin{equation}
\boldsymbol{Y} = \mathrm{FNN}(\mathrm{MultiHead}(\boldsymbol{H^P},\boldsymbol{H^D},\boldsymbol{H^0})).
\end{equation}

In each layer of the attention structure, $\boldsymbol{Y}$ is computed by interacting with both the feature vector $\boldsymbol{H^P}$ and the output $\boldsymbol{H^D}$ from the previous layer of the Decoder module. The resulting representation is then processed through a linear prediction layer in the final layer, ultimately generating the communication text output through a sampling procedure.

To train a model that enhances consistency in understanding and minimizes contradictions during inference, we can employ a likelihood-based training approach that differentiates between positive and negative data.

For positive data, the standard maximum likelihood loss is still employed during training, namely,
\begin{equation}
{L_{{D^ + }}} =  - \sum\limits_{i = 1}^{\left| R \right|} {\log \left( {P\left( {{r_i}\left| {{P^ - },} \theta ,\right.R, R < i} \right)} \right)} ,
\end{equation}
where ${P^ - }$ and $R$  correspond to  $\boldsymbol{H^P} $ and  $\boldsymbol{H^D} $ in the communication text task, respectively, and ${r_i} $ is converted from true semantics to the communication text output by the decoder end.

For negative data, likelihood training is used to minimize the possibility of contradictions:
\begin{equation}
{L_{{D^ - }}} =  - \sum\limits_{i = 1}^{\left| R \right|} {\log \left( {1 - P\left( {{r_i}\left| {{P^ - },} \theta ,\right.R,R < i} \right)} \right)} ,
\end{equation}
where ${P^-}$and $R$ are the premises and inference in the inference dataset, respectively, and the output target of ${r_i} $  is also the inference data.

According to (37), the joint loss training strategy helps the model maintain consistency and penalize contradictions. This approach is particularly well-suited for semantic feature-guided text generation, as it separates the consistency checking from the generation process. Consequently, the model generates text that is consistent with the input features via an enhancement step,
\begin{equation}
{L_D} \leftarrow \gamma   {L_{{D^ + }}} + (1 - \gamma   ){L_{{D^ - }}},
\end{equation}
where $\gamma  $ is set to 0.1. 
\section{Simulation Verification}
\subsection{Experimental Settings}


An experimental assessment was conducted to evaluate the effectiveness of the proposed semantic security approach for wireless network communication data, utilizing a knowledge graph-based methodology. The dataset used in this study is the WebNLG dataset \cite{10613766}, which includes a wide range of knowledge graphs and associated texts, spanning domains such as airports, artists, cities, and more. The communication dataset consists of 1,000 sentences, with a transmission interval of 0.37 seconds and a bandwidth of 45 kHz. Simulation platforms use signal sources, noise generators, and SNR parameters to model different signal-to-noise ratio levels. Five distinct SNR values are selected to simulate five different channel conditions, each representing a unique signal-to-noise ratio scenario. In the decryption bit-length experiments, the channel conditions are assumed as shown in Table~\ref{tab3}. Conduct 100 simulation experiments and calculate the average value. In the other experiments, the SNR ranges from 5 dB to 30 dB.

\begin{table}[!t] 
\caption{Parameters of the transmission channel\label{tab3}}
\centering
\begin{tabular}{|c|c|}
\hline
\textbf{Channel number}	& \textbf{Signal-to-Noise Ratio/dB}	\\
\hline
Channel 1	&5	 \\
\hline
Channel 2 		& 10\\	
\hline
Channel 3 		& 15\\	  
\hline
Channel 4 		& 20\\
\hline
Channel 5 		& 25\\
\hline
\end{tabular}
\end{table}

For comparison, a comparative analysis is conducted between (1) the hash-based semantic security communication method described in reference [8], (2) the NN-driven semantic security communication approach detailed in reference [9], and (3) the Deep Joint Source-Channel and Encryption Coding (DeepJSCEC)-based semantic secure communication method introduced in reference [10]. This comparison aims to provide a comprehensive evaluation of the secure transmission capabilities of the novel method proposed in this study.

(1) Decryption Bit Length: The decryption bit length is a critical factor in assessing the security effectiveness of cryptographic algorithms. It refers to the size of the cryptographic key employed in the encryption process, which directly influences the difficulty of adversaries attempting decryption. A longer key length increases the complexity of the decryption process, thereby strengthening the security of encrypted communications. The decryption bit length ($L_{\epsilon }$) in our simulations is rigorously determined using the following calculation formula.
\begin{equation}
L_{\epsilon } =\max\left ( \left [ \log_2(\mathcal{S}) \right ] , \Delta +\left [\log_2(B)  \right ]  \right ), 
\end{equation} 
Where $\mathcal{S}$ represents the keyspace cardinality (for symmetric encryption), 
$\Delta$  denotes the noise upper bound, and for Gaussian noise, 
$B=6\sigma$, where  $\sigma$ is the standard deviation of the noise distribution. The  parameter 
$\Delta$ , chosen as 128 bits, ensures that the decryption failure probability remains below $2^{-128}$.

(2) DFP: This refers to the probability that an eavesdropper fails to detect transmission activity during multiple instances within the data transmission period. A higher DFP indicates a greater safety margin. As such, this metric serves as an indirect measure of the security of wireless communication data. By conducting DFP testing, the effectiveness of the semantic security method for wireless communication can be assessed. The formula for calculating the detection failure probability is as follows:
\begin{equation}
{P_{fail}} = 1 - DEP{^{\frac{{{\delta _{i}}}}{{{\delta _{all}}}} \times F}}, 
\end{equation}
where ${\delta _{i}}$ represents the amount of data transmitted, $\delta _{all}$ represents the covert rate, and $F$ denotes the detection frequency. The Detection Error Probability (DEP) refers to the likelihood of making an incorrect decision when attempting to detect wireless transmission activities.

(3) BLEU score: The BLEU score is a critical metric for evaluating how well the data, after undergoing processing, retains the semantic essence of its original form. It measures the semantic similarity between the transformed data and the original, with a higher BLEU score indicating that the processed data closely resembles the original in terms of semantic content. A high BLEU score suggests that the security method effectively preserves key semantic information while securing the data, ensuring that the data's security is maintained without compromising its semantic integrity.
\subsection{Experimental Results}

\begin{figure}[!t]
\centering
\subfloat{%
  \includegraphics[width=0.45\textwidth]{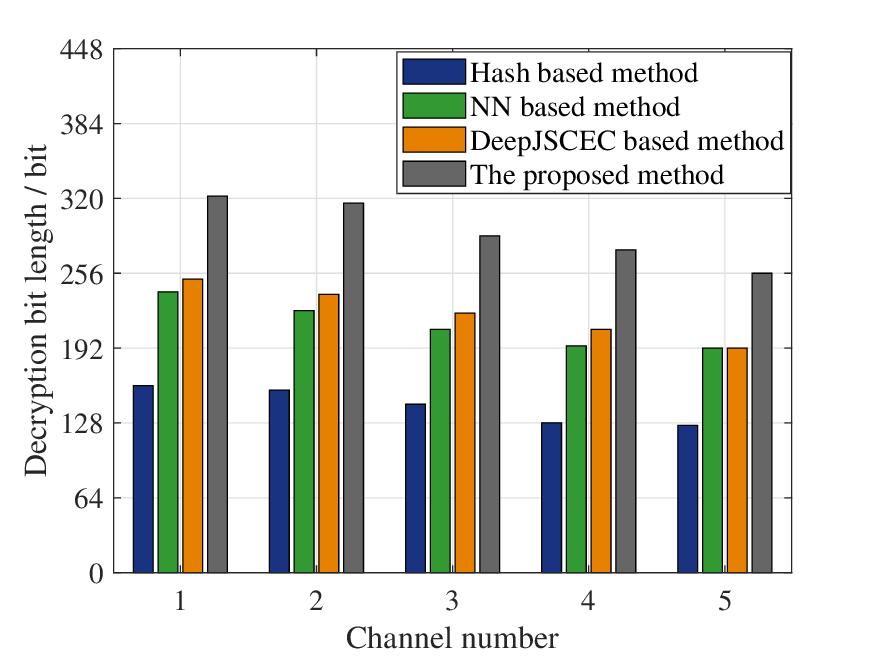}%
}
\caption{Decryption bit length result.\label{fig711}}
\end{figure}

Fig.~\ref{fig711} shows how the decryption bit length changes as different encryption methods are applied across five different SNRs. From Fig.~\ref{fig711}, we see that the proposed method achieves a significant improvement in decryption bit length compared to existing approaches. This is because our proposed encryption method has a larger key space. 
In Fig.~\ref{fig711}, we can also see that he proposed method consistently maintains high decryption bit-length values across all test conditions. This robustness stems from its adaptive design, which ensures strong security guarantees without significant computational overhead. This stems from the fact that the proposed encryption scheme alters the signal distribution through constellation diagonal encryption and MP-WFRFT, making it difficult for eavesdroppers to extract useful information from the received signal. This enhances the security of the communication system. 
Fig.~\ref{fig711} demonstrates that the proposed method effectively increases the decryption bit length, thereby enhancing the difficulty of decryption. The proposed approach outperforms the hash-based, neural network-based, and DeepJSCEC-based methods in terms of decryption bit length. Despite the increase in decryption bit length, the rise in latency remains moderate, thereby underscoring the proposed method's superior performance in delivering enhanced semantic security for wireless communication.

\begin{figure}[!t]
\centering
\subfloat{%
  \includegraphics[width=0.45\textwidth]{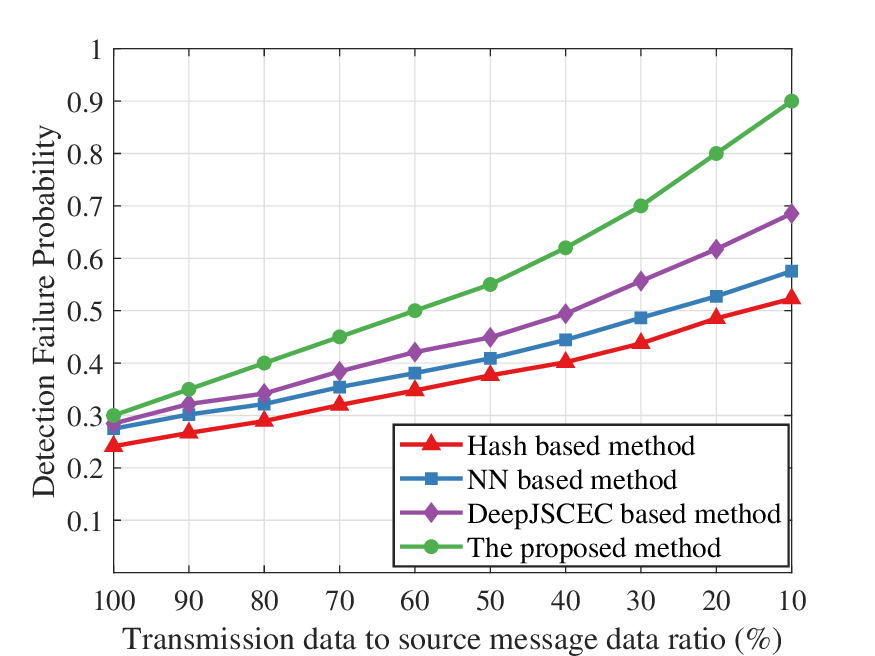}%
}
\caption{Detection Failure Probability results.\label{fig1033}}
\end{figure}

Fig.~\ref{fig1033} shows how detection failure probability 
 of the eavesdropper changes as the ratio of transmitted semantic information to source data volume decreases. In Fig.~\ref{fig1033}, we can see that, the proposed knowledge graph-based hybrid encryption method increases the detection failure probability of the eavesdropper by up to $29\%$, $55\%$, and $69\%$  compared to the DeepJSCEC-based, neural network-based, and hash-based encryption methods, respectively. The $29\%$ gain stems from the higher compression ratio of the proposed knowledge graph-based hybrid encryption method, compared to the DeepJSCEC-based encryption method, when applied to textual data. The $55\%$ gain stems from the fact that the proposed knowledge graph-based hybrid encryption approach transmits less data compared to the neural network-based encryption approach. The $69\%$ gain stems from the fact that the proposed knowledge graph-based hybrid encryption approach reduces the amount of data transmitted by constructing a knowledge graph through text processing, thereby increasing the detection failure probability.

\begin{figure}[!t]
\centering
\subfloat{%
  \includegraphics[width=0.45\textwidth]{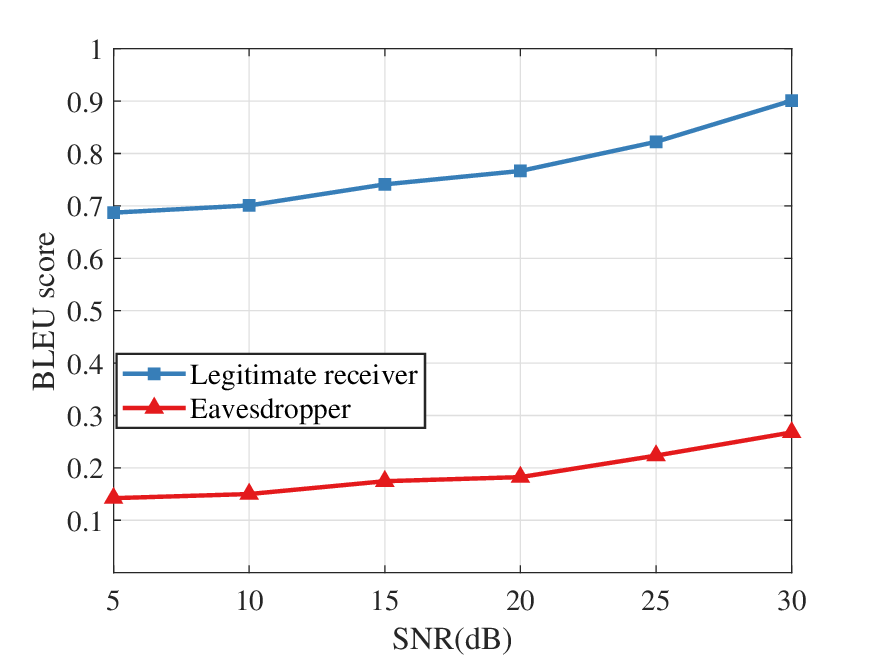}%
}
\centering
\caption{ BLEU score of legitimate receiver and eavesdropper. \label{fig8}}
\end{figure}

In Fig.~\ref{fig8}, we show how the BLEU scores of the legitimate receiver and the eavesdropper change as the Rayleigh channel SNR increases. The legitimate receiver achieves a higher BLEU score due to the shared semantic knowledge base between the transmitter and receiver. This shared knowledge allows the legitimate receiver to effectively decode and reconstruct the transmitted textual information accurately, even under varying channel conditions. Furthermore, Fig.~\ref{fig8} demonstrates that the proposed method effectively ensures data security, as the eavesdropper’s BLEU score remains below 0.3, while the legitimate receiver’s score stabilizes around 0.9. This indicates that the legitimate receiver can recover nearly the entire textual content, while the eavesdropper fails to reconstruct the original text. These results highlight the robustness of the proposed encryption method and its effectiveness in ensuring data security, as the eavesdropper cannot access the transmitted information, even under high SNR.

\begin{figure}[!t]
\centering
\subfloat{%
\includegraphics[width=0.45\textwidth]{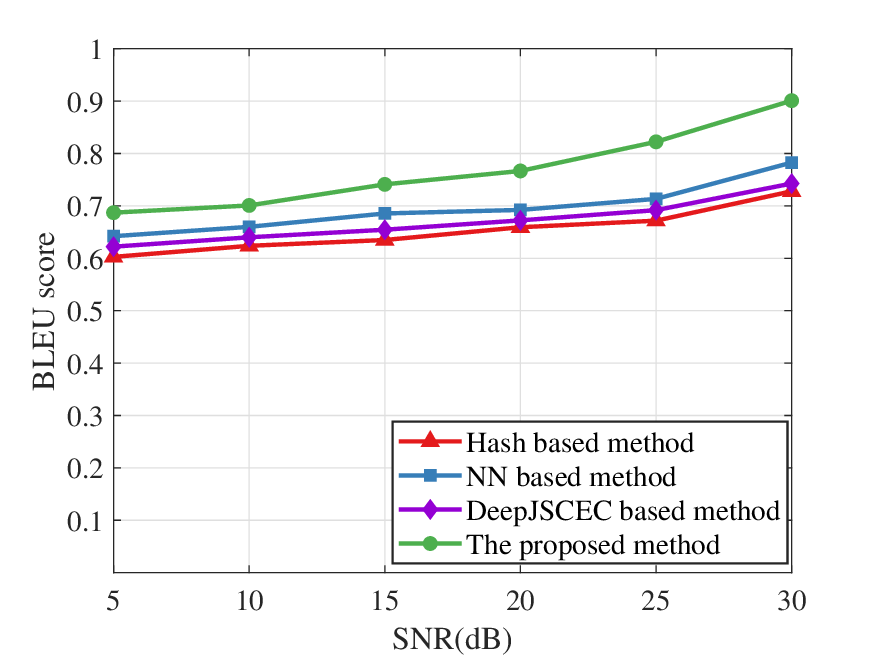}%
}
\centering
\caption{ BLEU scores for different methods.\label{fig311}}
\end{figure}

Fig.~\ref{fig311} shows the BLEU score as the Rayleigh channel SNR increases. This simulation is conducted to evaluate robustness of our approach in ensuring semantic security during transmission.
It can be observed that the  BLEU score improves with increasing SNR. This improvement is attributed to the fact that the proposed encryption algorithm provides richer contextual and semantic information, effectively reducing ambiguity and enhancing the accuracy and fluency of textual data, which leads to a significant increase in the BLEU score. 
Furthermore, Fig.~\ref{fig311} demonstrates that the proposed method maintains consistently high BLEU scores across most SNR ranges, with only minor fluctuations, thereby achieving better robustness.
This robustness is due to the encryption algorithm operating in the frequency domain, which ensures robust performance. Additionally, Fig.~\ref{fig311} highlights that as the SNR approaches 30 dB, the advantages of the proposed method become particularly evident. 
Even at lower SNR levels, ranging from 5 dB to 25 dB, our method exhibits robust performance, maintaining consistently high BLEU scores. These results demonstrate that the proposed encryption algorithm effectively preserves semantic integrity, thereby enhancing transmission accuracy under poor channel conditions.

\section{ Conclusion }

In this article, we present a novel framework for the secure transmission of textual data in a semantic communication system based on knowledge graphs. We consider a communication system that seeks to maximize the semantic accuracy of text recovery for legitimate users while minimizing that of the eavesdropper. To address this problem, we transform the text data into knowledge graphs to reduce the volume of data transmission and combine constellation diagonal encryption with MP-WFRFT encryption for channel encryption. On the receiver side, legitimate users who share a common knowledge base with the transmitter can accurately decrypt and recover the original text. In contrast, eavesdroppers, lacking both the private keys and access to the knowledge base, face significant difficulty in decrypting and recovering the textual data. Numerical evaluations on real-world datasets demonstrate that, compared to three representative existing methods—the hash-based method, the NN-based method and the DeepJSCEC-based method in semantic communication, the proposed method significantly enhances text transmission accuracy and security performance.
\bibliographystyle{IEEEtran}
\bibliography{IEEEabrv,ref}

\end{document}